\documentclass[superscriptaddress,aps,pra,twocolumn,showpacs,nofootinbib,longbibliography]{revtex4-2}
\usepackage{amsmath,amssymb,amsthm}
\usepackage{easybmat}
\usepackage[colorlinks=true,citecolor=blue,urlcolor=blue, linkcolor=magenta]{hyperref}
\usepackage[pdftex]{graphicx}
\usepackage{times,txfonts}
\usepackage{braket}
\usepackage{color}
\usepackage{natbib}
\newcommand{\be}{\begin{equation}}
	\newcommand{\ee}{\end{equation}}
\newcommand{\ba}{\begin{eqnarray}}
	\newcommand{\ea}{\end{eqnarray}}

\usepackage{physics}
\usepackage{yfonts}
\usepackage{graphicx} 
\usepackage{amsmath}
\usepackage{mathtools}

\begin{document}
\title{Bipartite OTOC in open quantum systems: information scrambling and irreversibility}

\author{Baibhab Bose\textsuperscript{}}
\thanks{Corresponding author}
\email{baibhab.1@iitj.ac.in}
\author{Devvrat Tiwari\textsuperscript{}}
\email{devvrat.1@iitj.ac.in}
\author{Subhashish Banerjee\textsuperscript{}}
\email{subhashish@iitj.ac.in}
\affiliation{Indian Institute of Technology Jodhpur-342030, India\textsuperscript{}}

\date{\today}
\begin{abstract}
	The field of information scrambling has seen significant growth over the last decade, where the out-of-time-ordered correlator (OTOC) has emerged as a prominent tool to probe it. In this work, we use bipartite OTOC, a particular form of OTOC, to study information scrambling in the atom-field interaction models and the model of the Ising spin chain interacting with a tilted magnetic field. This is done considering the effects of open quantum systems. A relationship between information scrambling, using bipartite OTOC, and irreversibility, using entropy production, is probed under unitary dynamics. The equivalence of bipartite OTOC with operator entanglement is explicitly shown for the Ising model.
\end{abstract}

\maketitle


\date{January 2024}

\section{Introduction}
The relationship between the classical and quantum dynamics of chaotic systems is at the heart of the study of quantum chaos. Recently, there has been a resurgence of interest in many-body quantum chaos and its relationship to information scrambling. Information scrambling refers to the spreading of quantum information over a complex quantum system such that it is not accessible to simple probes. Interest in this field was rejuvenated recently by works on black hole information paradox~\cite{Maldacena2016, Hayden_2007}, and the problem of quantum thermalization~\cite{Srednicki_1994}. Traditionally, the ideas of random matrix theory were employed to probe quantum chaos~\cite{mehta2004random,wigner}. In recent times, the out-of-time-ordered correlator (OTOC) has emerged as a prominent tool to investigate information scrambling. This was first brought out in the context of superconductivity~\cite{Larkin1969QuasiclassicalMI} and was popularized by~\cite{Shenker2014}. The growth rate of the OTOC is commonly attributed to the classical Lyapunov exponent \cite{Hirscheatomfield}. OTOCs have been extensively used in the study of a number of systems, such as in quantum field theories~\cite{Standford_2015}, random unitary models~\cite{David_2018}, spin chains~\cite{Zhang_2019, Fortes_2020}, and quantum optical models~\cite{Hirscheatomfield, devvrat_OTOC, Mahaveer}, among others~\cite{García-Mata:2023, Swingle_2024, Swingle2018}. Various forms of the OTOC, such as the regularized and the physical OTOC, were also used in the context of the fluctuation-dissipation theorem, generalized to the case of the OTOC~\cite{Uedaflucdiss}.

Numerous qualitative aspects of the OTOC remain unaffected by the particular operators selected, provided that their locality remains constant. Thus, OTOCs averaged over (suitably distributed) random operators represent a significant simplification. It is possible to perform the uniform average over pairs of random unitary operators, supported across both sides of the bipartition of the system Hilbert space \cite{zanardibipotoc2}. The operational importance of this averaged bipartite OTOC is that it quantifies the scrambling of information at the quantum channel level, as well as the operator entanglement of the dynamics~\cite{zanopen44, Zanardi2001opent}. Further, Haar averaging over all the unitary matrices in the given Hilbert space eliminates any specific choice of unitary matrices to probe the information scrambling in the system. Thus, the averaged bipartite OTOC is an effective instrument for examining chaos and information scrambling in many-body quantum systems using unitary operators. This framework, expanded to include open quantum systems~\cite{Zanardiopenbipotoc1}, that is, the quantum systems impacted by their ambient environment, will be used in this paper.

The theory of open quantum systems provides a framework to study quantum systems affected by their surroundings \cite{Weiss2011,Breuer2007,Banerjee2018}. There has been significant growth in the applications of this theory in the field of quantum information~\cite{Omkar2016, Javid_2018, Vacchini_2011, Tiwari_2023}. The dynamics of an open quantum system can be understood using Markov and rotating wave approximations, characterized by the Gorini-Kossakowski-Lindblad-Sudarshan (GKLS) master equation \cite{GKLSpaper, Lindblad1976}. With advancements in theory and technology, the understanding of open systems has expanded from the limitations of Markovian dynamics to the domain of non-Markovianity \cite{Hall_2014, Rivas_2014, RevModPhys.88.021002, CHRUSCINSKI20221, banerjeepetrucione, vega_alonso, Utagi2020}. In a number of scenarios, the effect of open quantum systems on information scrambling have been realized \cite{zanopen21, Zhang_2019, zanopen31}. Further, the concept of information scrambling shares an information-theoretic connection to mutual information~\cite{touil2024information}, which is a key ingredient in the formulation of entropy production, characterizing irreversibility in the system~\cite{Landientropyprod, EspositoEntprodentcorre}. 

\begin{figure}
    \centering
    \includegraphics[width = 1\columnwidth]{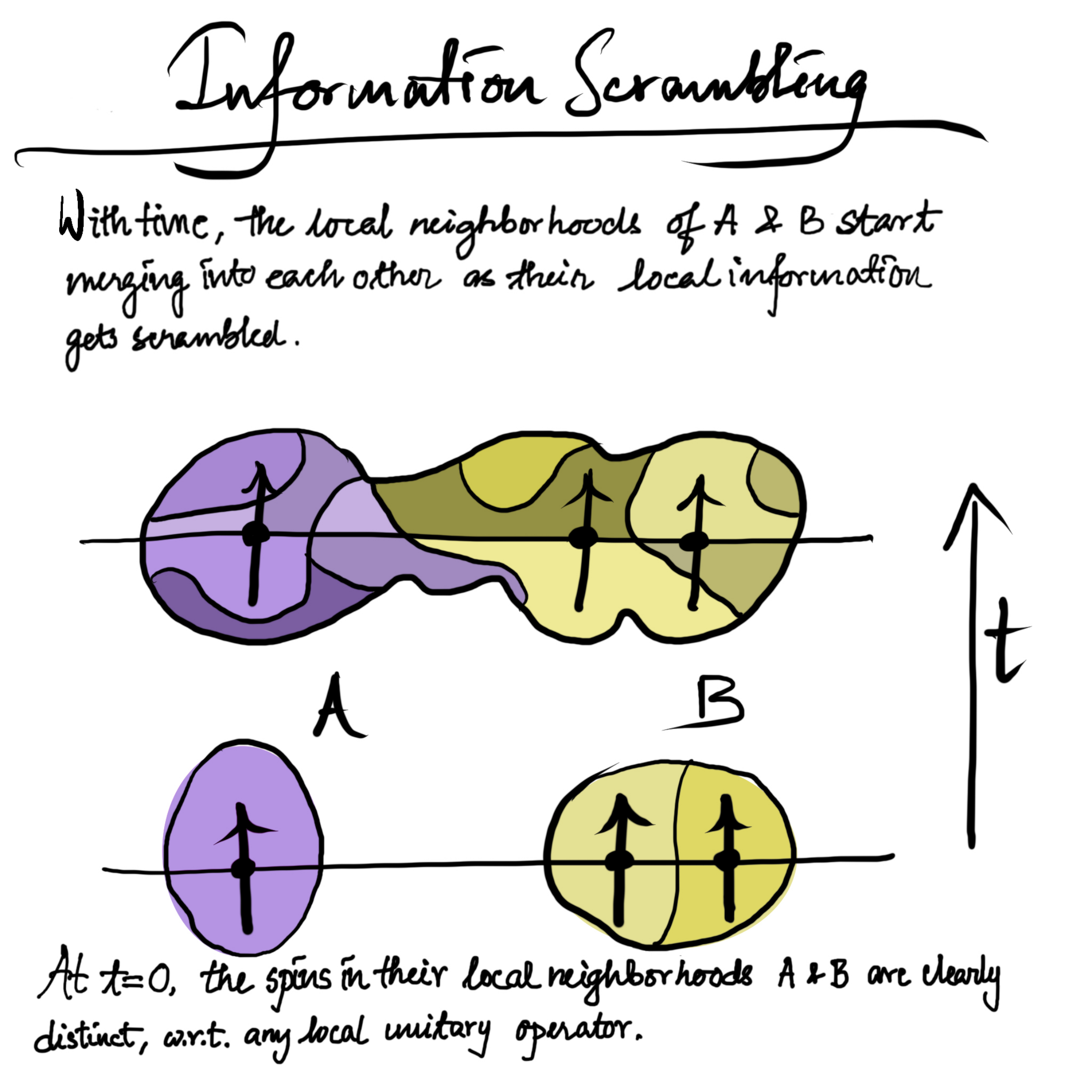}
    \caption{A toy model of an Ising spin chain partitioned into two subsystems with 1 spin in subsystem $A$ and the rest in subsystem $B$. The figure shows how, with time, the initial distinguishability of $A$ and $B$, due to the commutativity of the local operators, gets disrupted, and the bipartition loses its distinction.}
    \label{fig:Information_Scrambling_cartoon}
\end{figure}

Information scrambling invokes the scenario wherein there is a loss of distinguishability between two local neighborhoods, say $A$ and $B$, see Fig.~\ref{fig:Information_Scrambling_cartoon}. This causes irreversibility emerging from discarding any information contained locally, say in the state of $B$, wherein the non-local information shared between $A$ and $B$ becomes essential. This irreversibility can be probed by entropy production and hence motivates a relationship between information scrambling and entropy production. Entropy production takes center stage in the study of quantum thermodynamics~\cite{Landientropyprod, EspositoEntprodentcorre}, which is one of the prominent applications, in recent times, of the theory of open quantum systems, fuelled by the development of quantum technologies and devices~\cite{Gemmer2009, Binder_book, sai_janet_book}. The study of quantum thermodynamics in the presence of system-bath coupling in the (non-)Markovian regime is a challenging task~\cite{thomas_SB, kosloff_2022, Alicki_battery, tiwari2024strong}.

The aim of this work is to study the information scrambling in the form of OTOC, in particular bipartite OTOC, in various models. This includes models of atom-field interaction, particularly the Dicke and Tavis-Cummings (TC) models and the Ising model with a tilted magnetic field. These models have been shown to exhibit quantum chaotic properties in particular parameter regimes~\cite{Hirscheatomfield, devvrat_OTOC, TiltedMfieldofarul}. This is done in the presence of open quantum system effects. Further, a relationship between information scrambling and irreversibility (using entropy production) is investigated. 

The plan of the paper is as follows. In Sec.~\ref{sec_bip_OTOC}, we briefly discuss the concept of bipartite OTOC and its form in the presence of open quantum system effects. Section~\ref{sec_dynamics_bip_OTOC} explores the dynamics of the bipartite OTOC for the Dicke, TC, and Ising models using the adjoint form of the GKLS master equation. A discussion on the relationship between the bipartite OTOC, operator entanglement, and entropy production is done in Sec.~\ref{sec_bip_OTOC_op_ent_ent_prod}, followed by conclusions in Sec.~\ref{sec_conclusion}.

\section{Bipartite OTOC}\label{sec_bip_OTOC}

The strength of the information scrambling in a quantum system under the Hamiltonian dynamics can be understood as the evolution of non-commutativity of local operators $V(t)=U^{\dagger}VU$ evolved in time with another operator $W$ at time $t=0$ and is given by the function
\begin{equation}
    C_{V,W}(t) \coloneqq \frac{1}{2} {\rm Tr}([V(t),W]^\dagger [V(t),W]\rho), \label{eq:1}
\end{equation}
where the average is taken using the maximally mixed state $\rho = \frac{\mathbb{I}}d$. 
The exponential nature of $C_{V, W}(t)$ in Eq. (\ref{eq:1}) indicates the chaotic nature of a quantum system. This has been verified in systems with semiclassical limits or systems with a large number of degrees of freedom. Quantum chaos, generally, refers to the exponentially fast dispersion of information throughout a
quantum many-body system. 
Mathematically, the `out-of-time-ordered correlator' (OTOC) corresponds to 
\begin{equation}\label{eq:2}
    F_{V,W}(t) \coloneqq  \left\langle V^{\dagger}(t)W^{\dagger}V(t)W\right\rangle,
\end{equation}
which is a four-point correlation function where the operators are out-of-time-ordered, and the average is taken with respect to the state $\rho$. The quantities in Eqs. (\ref{eq:1}) and (\ref{eq:2}) can be linked for the case where the operators $V(t)$ and $W$ are unitary via the relation
\begin{equation}\label{eq:3}
C_{V,W}(t)=1 - \Re\left\{F_{W,V}(t)\right\}. 
\end{equation}
The bipartite OTOC is defined over a Hilbert space that has two partitions and measures the evolution of the non-commutativity of two unitary operators from those two partitions. 
For OTOCs involving unitary operators, the choice of those operators is crucial. The conception of bipartite OTOC aims to get rid of this dependence on the choice of unitary operators. Since all the unitary operators are not equally sensitive to chaotic behavior, averaging over all possible random unitary operators renders a more general form of OTOC \cite{22nofromzanardipaper,30nofromzanardipaper,31nofromzanardi,32nozanardipaper,33nofromzanardipaper,34fromzanardipaper,35nofromzanardipaper}
We consider a finite-dimensional Hilbert space with a bipartition, such that
the composite Hilbert space is $\mathcal{H}=\mathcal{H}_A \otimes \mathcal{H}_B$. The function $C_{V_A,{W_B}}(t)$, Eq. \eqref{eq:3}, is Haar averaged over all independent unitary operators $V_A$ and $W_B$, which belong to subsystems A and B, respectively. This Haar averaged quantity is
\begin{equation}\label{eq:1st average}
    G(t)\coloneqq 1-\frac{1}{d}{\rm Re}\int{dVdW{\rm Tr}\left[V^{\dagger}_A(t) W^{\dagger}_B V_A(t) W_B\right]}.
\end{equation}
We now introduce another replica of the Hilbert space of $\mathcal{H}$, which is, $\mathcal{H}^{\prime}=\mathcal{H}_{A^{\prime}} \otimes \mathcal{H}_{B^{\prime}}$ with the same dimension as that of $\mathcal{H}$, i.e., $d={\rm dim}(\mathcal{H})=d_A d_B$ and $U_t=e^{-iHt}$ is the unitary evolution of the composite system $(A+B)$.
Invoking the swap operators $S_{AA^{\prime}}$ and $S_{BB'}$  (see Appendix A), which belongs to the Hilbert space of $\mathcal{H} \otimes \mathcal{H^{\prime}}$, the expression for the bipartite OTOC $G(t)$ is given by  \cite{Zanardiopenbipotoc1}
\begin{align}\label{Bip otoc unitary}
         G(t)&=1-\frac{1}{d}{\rm ReTr}\left(SU_t^{\dagger\otimes2} \frac{S_{AA^{\prime}}}{d_A} U_t^{\otimes2} \frac{S_{BB^{\prime}}}{d_B}\right). 
    \end{align}
The $S_{BB^{\prime}}$ can be replaced using the identities
\begin{equation}
       S=S_{AA^{\prime}}S_{BB^{\prime}}=S_{BB^{\prime}}S_{AA^{\prime}}
\end{equation}
and the fact that $S^2=S_{AA^{\prime}}^2=S_{BB^{\prime}}^2=\mathbb{I}$, which is natural for swap operators. Replacing $S_{BB^{\prime}}$ yields the final expression for the bipartite OTOC of a unitary evolution $U(t)$, which is given by
\begin{equation}\label{eq:5}
    G(t)=1-\frac{1}{d^2}{\rm Tr}(S_{AA^{\prime}}U_t^{\otimes2}S_{AA^{\prime}}U_t^{\dagger \otimes2}),
\end{equation}
the same expression is valid for $S_{BB^{\prime}}$ also.

\subsection{Bipartite OTOC for open quantum systems}
The evolution of open quantum systems involves a completely positive trace preserving (CPTP) quantum channel $\mathcal{E}$, which maps the initial system density matrix to a density matrix at a later time. To modify the bipartite OTOC for open systems, the unitary map must be replaced by a CPTP map  $\mathcal{E}$. In the Heisenberg picture, we employ the adjoint master equation, which deals with the time evolution of a certain operator in the system's Hilbert space. The CPTP map that acts on the operator is $\mathcal{E}^{\dagger}$, and the evolved operator is $\mathcal{E}^{\dagger}(V)$. The bipartite OTOC for the open system in terms of the Hilbert-Schmidt norm can then be written as \cite{Zanardiopenbipotoc1}
    \begin{align}
        G(\mathcal{E}^{\dagger})&\coloneqq \frac{1}{2d} \mathbb{E}_{V_A,V_B}[C_{V_A,W_B}(t)] =\frac{1}{2d} \mathbb{E}_{V_A,V_B}||[\mathcal{E}^{\dagger}(V_A),W_B]||^2_2 \nonumber \\
        &=\frac{1}{2d} \mathbb{E}_{V_A,V_B} {\rm Tr} \{ [\mathcal{E}^{\dagger}(V_A),W_B]^{\dagger}[\mathcal{E}^{\dagger}(V_A),W_B]\} , 
        \label{channeltr}
    \end{align}
where $V_A=V \otimes I_B, W_B=I_A\otimes W$ with $V \in \mathcal{U}(\mathcal{H}_A)$ and $W \in \mathcal{U}(\mathcal{H}_B)$, i.e., $V_A$ and $W_B$ belong to the set of unitary operators in the Hilbert space $\mathcal{H}_A$ and $\mathcal{H}_B$, respectively. $\mathbb{E}_{V_A, W_B}$ denotes the Haar averaging with respect to all possible unitary operators in the Hilbert space, that is, 
\begin{equation}
    \mathbb{E}_{X\in \mathcal{G}}({\rm ...})\equiv \int_{X\sim Haar}dX({\rm ...}).
\end{equation}
Making use of the swap operators $S_{AA'}$ and $S$, the bipartite OTOC for open quantum systems can be expressed as 
\begin{align}
    G(\mathcal{E}^{\dagger})=&\frac{1}{d} \bigg\{ {\rm Tr}\big( S\mathcal{E}^{\dagger \otimes 2} \frac{S_{AA^{\prime}}}{d_A} \big)- \frac{1}{d}{\rm Tr} \big( S_{AA^{\prime}} \mathcal{E}^{\dagger \otimes 2} S_{AA^{\prime}} \big) \bigg\}, \nonumber \\
   =&\frac{1}{d^2}{\rm Tr} \{ (Sd_B-S_{AA^{\prime}}) \mathcal{E}^{\dagger \otimes 2} S_{AA^{\prime}} \}.
   \label{eq_avg_bip_otoc_open}
\end{align}
Hence, if we know the form of swap operators in the Hilbert space and the form of the quantum channel, we can calculate the bipartite OTOC for the open quantum system. 
The CPTP adjoint map affects only the operator belonging to the A subsystem. The effect of the map is trivial over the operator from the B subsystem, although the adjoint map acts on the operator that spans the full subspace. Due to the dependence of the bipartite OTOC on the swap operator acting on both A and B subspaces, scrambling of local information between them is possible for evolution under both unitary and adjoint maps.
In the following sections, we explore the dynamics of the bipartite OTOC for the models that are known to have a chaotic regime. 

\section{Dynamics of Bipartite OTOC for various models}\label{sec_dynamics_bip_OTOC}
Here, the bipartite OTOC is calculated for three models, namely the Dicke and the Tavis-Cummings (TC) models and the Ising model with a tilted magnetic field. For the first two cases, the bipartition is composed of the spin and radiation parts, which are quantum optical models of atom-field interaction~\cite{Larson_Dicke, devvrat_OTOC}. For the Ising model in the tilted magnetic field, the bipartition is that of one spin and all other spins. For all the models, open-system effects will be considered.

\subsection{Dicke model}
Here, we consider the $N$-qubit Dicke model consisting of $N$ two-level atoms with transition frequencies $\omega_0$, coupled to a single mode of a quantized radiation field of frequency $\omega_c$~\cite{RH_Dicke, Kirton_review}. The Hamiltonian ($\hbar = 1$) of the system is given by
\begin{align}
    H_{Dicke}=\omega_0 J_z +\omega_c a^{\dagger}a +\frac{\lambda}{\sqrt{N}} J_x(a+a^{\dagger}).
\end{align}
The collective angular momentum operators $J_i$ ($i=x, z$), in terms of a pseudospin of length $j=N/2$, describe the ensemble of two-level atoms. The single mode quantized radiation field is depicted by the number operator $a^\dagger a$, where $a$ and $a^\dagger$ are the bosonic creation and annihilation operators, respectively. The third term gives the coupling between the quantized radiation field and the atoms, where $\lambda$ denotes the interaction strength and $\sqrt{N}$ is the scaling factor.

\begin{figure}
    \centering
    \includegraphics[width = 1\columnwidth]{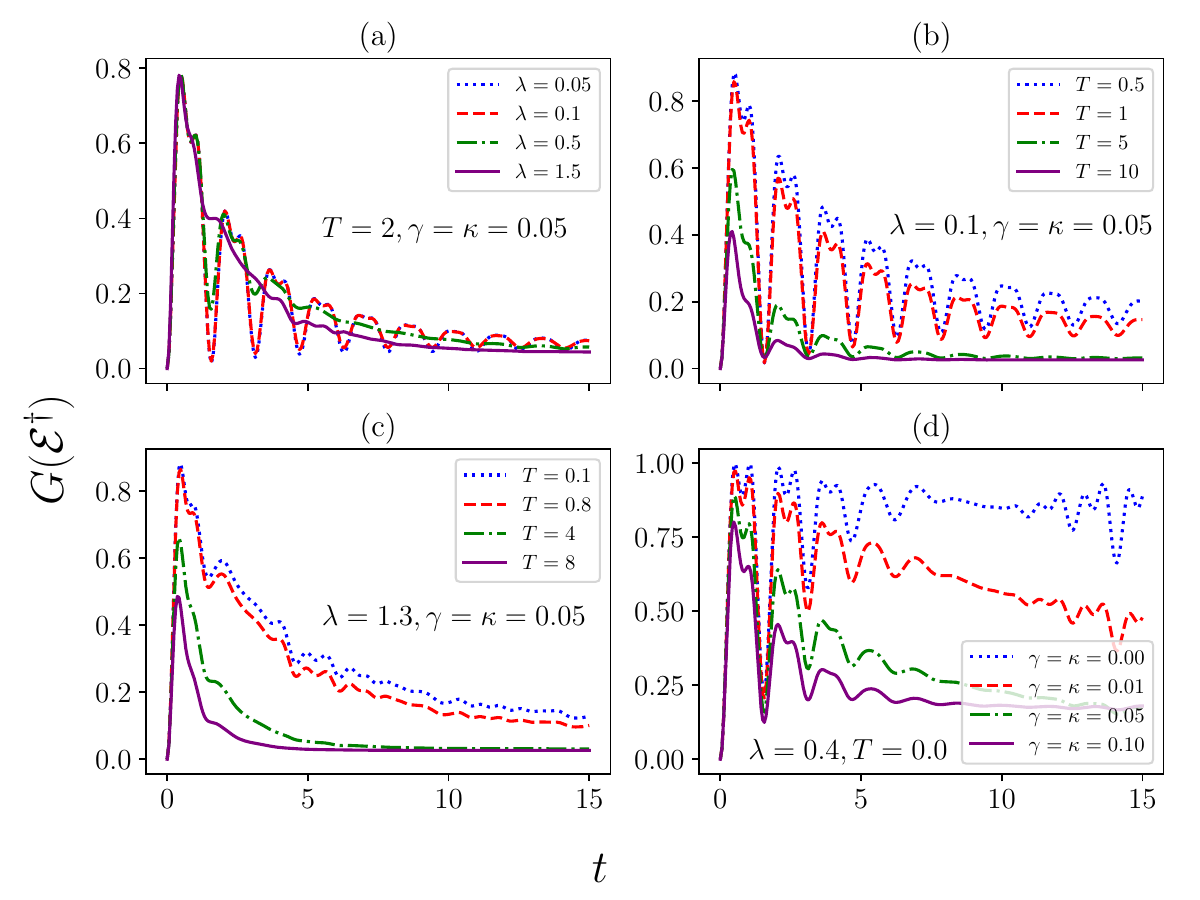}
    \caption{Variation of averaged bipartite OTOC $G(\mathcal{E}^\dagger)$ for the Dicke model with time $t$ for (a) different values of spin-cavity field interaction strength $\lambda$, (b) different values of temperature $T$ at $\lambda = 0.1$ and (c) at $\lambda = 1.3$, and (d) different values of dissipative strengths $\gamma$ and $\kappa$. The value of parameters are: $\omega_0 = 2$, $\omega_c = 2$.}
    \label{fig:varied_lambda_bip_OTOC_dickeNw}
\end{figure}

The two sources of dissipation in this atom-field system can be attributed to spontaneous emission and cavity decay. To this end, we need to consider the above model as an open quantum system, with its dynamics being governed by a Gorini-Kossakowski-Sudarshan-Lindblad (GKSL) master equation. The adjoint form of the GKSL master equation, which is required to study the evolution of operators, is given below
\begin{align}\label{MasterEqu}
    \frac{\partial}{\partial t} \mathcal{O} = \mathcal{L}^\dagger(\mathcal{O}) &= i[H_{Dicke},\mathcal{O}] + \gamma (N_{thA}+1) \left[ J_+ \mathcal{O} J_- -\frac{1}{2}\left\{\mathcal{O}, J_+ J_- \right\} \right] \nonumber \\
    &+ \gamma N_{thA} \left[J_- \mathcal{O} J_+ -\frac{1}{2}\left\{\mathcal{O}, J_- J_+ \right\} \right] \nonumber \\
    &+ \kappa(N_{thB}+1) \left[a^{\dagger} \mathcal{O} a  -\frac{1}{2}\left\{\mathcal{O}, a^{\dagger}a \right\} \right] \nonumber \\
    &+ \kappa N_{thB} \left[a \mathcal{O} a^{\dagger}  -\frac{1}{2}\left\{\mathcal{O}, aa^{\dagger} \right\} \right],
\end{align}%
where $\gamma$ acts as a dissipating factor modelling spontaneous emission, and $\kappa$ is responsible for cavity decay. Further, $N_{th_k} = \frac{1}{e^{\beta\omega_k} - 1}$ with $\beta = \frac{1}{k_B T}$, and $k = A$, $B$ denoting the subspaces $A$ and $B$, respectively. 
By vectorizing this master equation, the Lindbladian superoperator $\mathcal{L}^\dagger$ and, in turn, the CPTP map $\mathcal{E}^{\dagger} = e^{t\mathcal{L}^\dagger}$ is obtained. We consider the atomic system to belong to the $A$ subspace ($\omega_A = \omega_0$), and the radiation field is relegated to the $B$ subspace ($\omega_B = \omega_C$).
By applying the swap operators for the specific dimensions, the bipartite OTOC is obtained. Further, the Fock space of the field part is truncated in such a manner that increasing the dimension of the Fock space after a point doesn't affect the dynamics ~\cite{devvrat_OTOC, Hirsch_2011}. 
The dynamics of the bipartite OTOC for the Dicke model for different values of atom-field interaction strength $\lambda$ is shown in Fig.~\ref{fig:varied_lambda_bip_OTOC_dickeNw}(a).
We observe that as the atom-field coupling strength $\lambda$ increases, the fluctuations (or revivals) in the Bipartite OTOC $G(\mathcal{E}^{\dagger})$ decrease; that is, the curves smooth out. This is predominant when the coupling parameter approaches the critical value $\lambda_c = \frac{\sqrt{\omega\omega_0}}{2}$, where the Dicke model transitions from the normal regime to the superradiant regime. In \cite{devvrat_OTOC}, an exponential behavior of OTOC was observed when the Dicke model underwent a phase transition from the normal to the superradiant regime. A consistent result is observed here for the bipartite OTOC, where the smoothening of the curve in the superradiant regime, characterized here by $\lambda = 1$, could be considered an indicator of the onset of quantum chaos. Further, in Figs.~\ref{fig:varied_lambda_bip_OTOC_dickeNw}(b) and~\ref{fig:varied_lambda_bip_OTOC_dickeNw}(c), we plot the variation of the bipartite OTOC for different temperatures for given values of $\lambda$. It is observed that the bipartite OTOC takes the highest values for zero temperature. As the temperature increases, the values of the bipartite OTOC reduce. Further, in Fig.~\ref{fig:varied_lambda_bip_OTOC_dickeNw}(d), we plot the variation of the bipartite OTOC with time for different values of the dissipative strength $\gamma$ and $\kappa$ at zero temperature. In this case, the bipartite OTOC reduces as we increase the dissipative strength. For $\gamma = \kappa = 0$, the bipartite OTOC does not dissipate with time. 

\subsection{Tavis-Cummings model}
The Tavis-Cummings (TC) model can be obtained from the Dicke model by removing the counter-rotating terms, and its Hamiltonian is
\begin{equation}
    H_{\text{TC}} = \omega_0  J_z + \omega_c a ^\dagger a + \frac{\lambda}{2\sqrt{N}} \left( J_+ a +  J_- a^\dagger \right).
\end{equation}
The total number of the excitations $\mathcal{N} = a^\dagger a + J_z$ in this model remains conserved.
The sources of dissipation in this model are the same as in the Dicke model. By replacing $H_{Dicke}$ in Eq.~\eqref{MasterEqu} by $H_{TC}$, we arrive at an adjoint master equation for the TC model. The swap operators for this model remain the same as those for the Dicke model.
The variation of the bipartite OTOC for the TC model is similar to the variation of the bipartite OTOC for the Dicke model. Like the Dicke model, as the coupling strength increases, the fluctuations (or revivals) in the bipartite OTOC $G(\mathcal{E}^{\dagger})$ decrease. 
The fluctuations in the bipartite OTOC act as an identifier of an integrable system; the smoothness of the bipartite OTOC indicates non-integrability. 

\subsection{Ising model with tilted magnetic field}
Next, we consider an Ising chain, which interacts with a tilted magnetic field. This model has been found to be useful for understanding the chaotic behavior of a quantum system and has been studied extensively from statistical physics and quantum information perspectives~\cite{TiltedMfieldofarul}. The Hamiltonian $H_{\theta}$ for this model is given by
\begin{align}
    H_{\theta}(J,B,\theta)= H_A+H_B+H_{AB},
\end{align}
    where,
\begin{align}\label{Ham_Ising}
    H_A &= \mathcal{B}\left(\sin(\theta) \sigma^x_1 + \cos(\theta) \sigma^z_1\right), \nonumber \\
    H_B &= \mathcal{B} \sum_{i=2}^N  \left(\sin(\theta) \sigma^x_i +\cos(\theta) \sigma^z_i\right), ~~~~\text{and} \nonumber \\
    H_{AB} &=J\sum_i^{N-1} \sigma_{i}^z \sigma_{j+1}^z,
\end{align}
where $\mathcal{B}$ denotes the magnetic field strength, $\theta$ is the angle of tilt of the magnetic field, varying which one can take the model from the integrable to the non-integrable regime, and $J$ is the spin-spin coupling strength. $H_A$ and $H_B$ defines the Hamiltonian for $A$ and $B$ bipartitions, respectively. $\sigma^z$ and $\sigma^x$ are the Pauli spin matrices. For $\theta = 0$, the magnetic field is longitudinal, where the model is almost trivially integrable, and the spectrum is highly degenerate. When $\theta=90^{\circ}$, the field is transverse, and the model is
still integrable because of the Jordan-Wigner transformation that maps the model to the noninteracting fermions. 
When $0<\theta<90^{\circ}$, the Jordan-Wigner transformation results in interacting fermions, making the model non-integrable~\cite{Osborne_2002}. 

\begin{figure}
    \centering
    \includegraphics[width = 1\columnwidth]{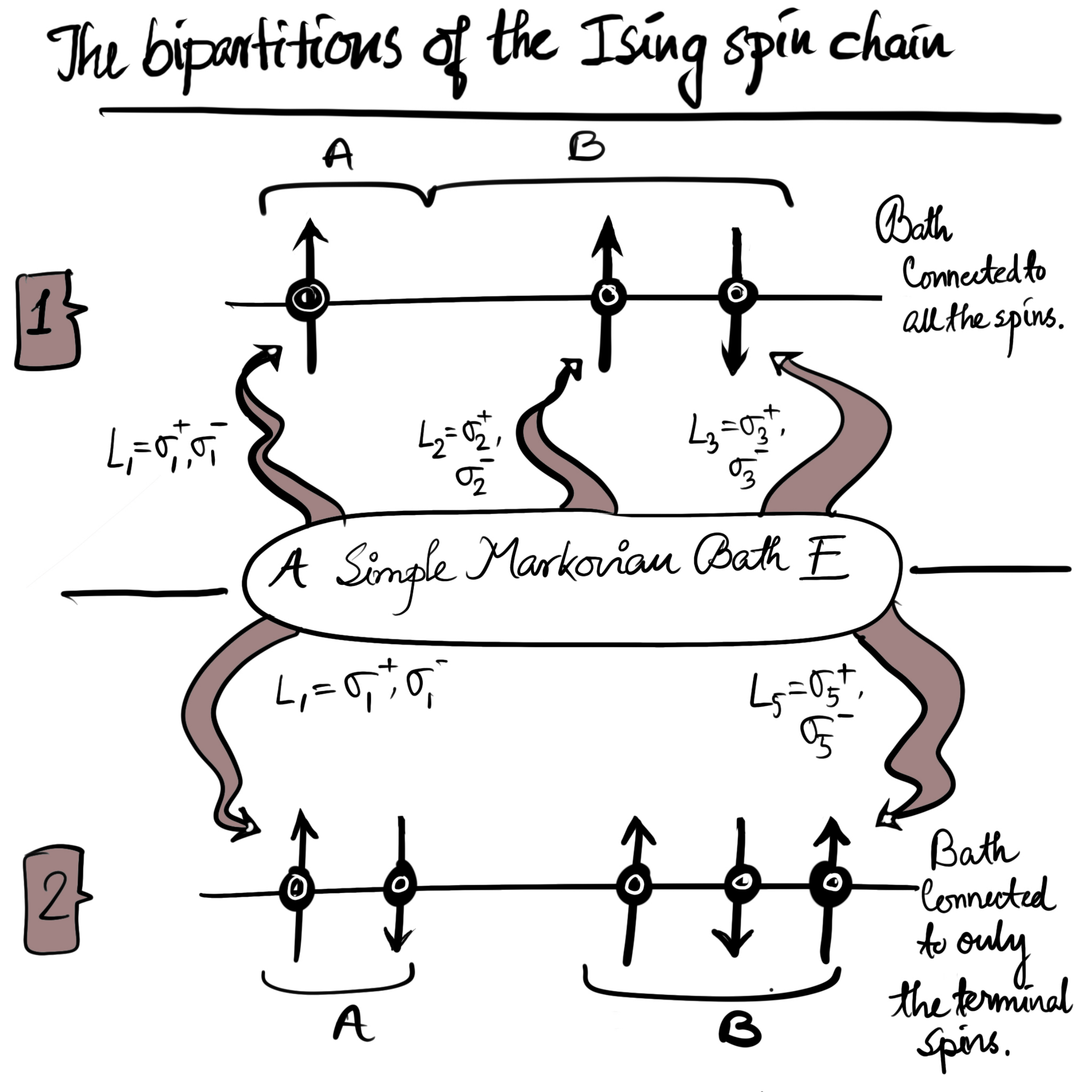}
    \caption{A schematic sketch illustrating the bipartitions and action of the bath in the Ising chain for calculating the bipartite OTOC.}
    \label{fig:bipartitions_sketch}
\end{figure}

In Fig.~\ref{fig:bipartitions_sketch}, we sketch the bipartition in the Ising model. In the first case, all the spins are acted upon by similar GKSL Lindbladians. In the other case, we consider a scenario where the first and the last spins are being acted upon by bosonic baths at different temperatures. We now take recourse to the Heisenberg formalism, where we use the adjoint form of the GKSL master equation to analyze the time evolution of a system operator $A$, which is given by
\begin{align}\label{master_eq_Ising}
    \frac{\partial}{\partial t} \mathcal{O} &=  i[H_{\theta},\mathcal{O}] + \sum_{j=1}^N \bigg( \gamma (N_{th_j}+1) \bigg[ \sigma_+ ^j \mathcal{O} \sigma_- ^j -\frac{1}{2}\{\mathcal{O}, \sigma_+ ^j \sigma_- ^j \} \bigg] \nonumber \\
    &+ \gamma N_{th_j} \bigg[\sigma_- ^j \mathcal{O} \sigma_+ ^j -\frac{1}{2}\{\mathcal{O}, \sigma_- ^j \sigma_+ ^j \} \bigg]  \bigg),
\end{align}%
where $N$ is the number of spins in the Ising chain being acted upon by the bath, and $\gamma$ is the dissipation factor. Further, $N_{th_j} = \frac{1}{e^{\beta \omega_j} - 1}$, where $\omega_j$ is the transition frequency of the spin at site $j$, which in this case, for each spin is $2\mathcal{B}$.

\begin{figure}
    \centering
    \includegraphics[width = 1\columnwidth]{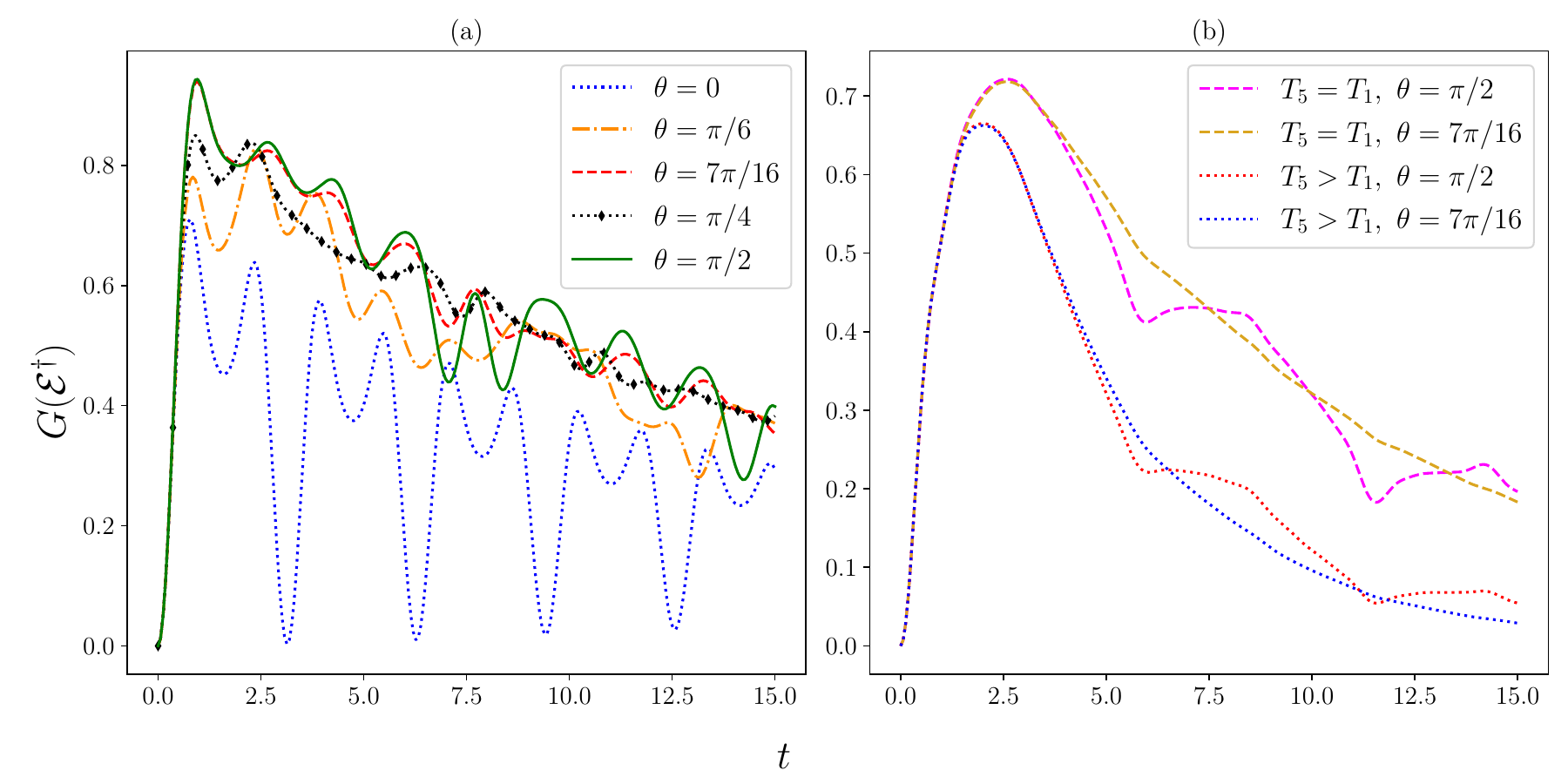}
    \caption{Variation of averaged bipartite OTOC for (a) the Ising model in the tilted magnetic field $G(\mathcal{E}^\dagger)$ with time $t$ for different values of angle of tilt of the magnetic field $\theta$, and (b) a system with two bipartitions with subsystem A having 2 spins and B having 3 spins with the first and last spins experiencing the influence of baths. Two cases are considered in (b), where the two baths are at the same ($T_1=T_5=1$) or different temperatures ($T_1=1$, and $T_5=5$). The values of parameters are (a): $T = 1$, $\gamma = 0.01, \mathcal{B}= J= 0.5$, (b): $J= \mathcal{B}=1,\gamma = 0.05$.}
    \label{fig:Theta_model_bip_OTOC_corrected}
\end{figure}

The variation of the bipartite OTOC for this model for four different angles of tilt, including the two extreme cases of $\theta=0^{\circ}$ and $\theta=90^{\circ}$ is plotted in Fig.~\ref{fig:Theta_model_bip_OTOC_corrected}(a).
The bipartition, in this case, is considered such that one subsystem has a single spin, and the remaining spins are in the other subsystem. Here, we observe that for a given coupling constant $J=0.5$, the fluctuations in the bipartite OTOC $ G(\mathcal{E}^{\dagger}) $ tend to smoothen out as the angle of the tilt of the magnetic field increases. However, the bipartite OTOC for angle $\theta = 0^\circ$ is oscillatory, consistent with the integrable behavior of the Ising model for this angle. 

Further, we study the bipartite OTOC in the above Ising model under the assumption that the bath acts only on the first and the last spins of the Ising chain. The bipartition of the system is taken as 2 spins in the $A$ subsystem and 3 spins in the $B$ subsystem, as illustrated in the second sketch of Fig.~\ref{fig:bipartitions_sketch}. Here, we can study a spin chain attached to the baths at the same or different temperatures on either side. In this case, only the first and last spins are subjected to the dissipating Lindblad operators $L_j$'s as $\sigma_{\pm}^j$ (where $j=1, N)$. To this end, the above master equation [Eq.~\eqref{master_eq_Ising}] with appropriate Lindblad operators is used. The bipartite OTOC for this system is shown in Fig.~\ref{fig:Theta_model_bip_OTOC_corrected}(b) for two different angles of tilt with the magnetic field for the same or different temperatures of the two baths.
It is evident from the figure that the bipartite OTOC takes lesser values in the presence of different bath temperatures, with the last spin having a higher temperature than the first one. Furthermore, a smooth graph of the bipartite OTOC is obtained for the angle $7\pi/16$, indicating non-integrable behavior, whereas the graph for the angle $\pi/2$ has fluctuations in it, indicative of integrability. This is consistent with the (non-)integrable behavior of the Ising model, Eq.~\eqref{Ham_Ising}. Interestingly, from the plots of the bipartite OTOC in Figs.~\ref{fig:Theta_model_bip_OTOC_corrected}(a) and (b), for the angle $\theta = \pi/2$, we observe that the detection of integrable behavior is hampered by the dissipation.

\section{Bipartite OTOC, operator entanglement, and entropy production}\label{sec_bip_OTOC_op_ent_ent_prod}
\begin{figure}
    \centering
    \includegraphics[width = 1\columnwidth]{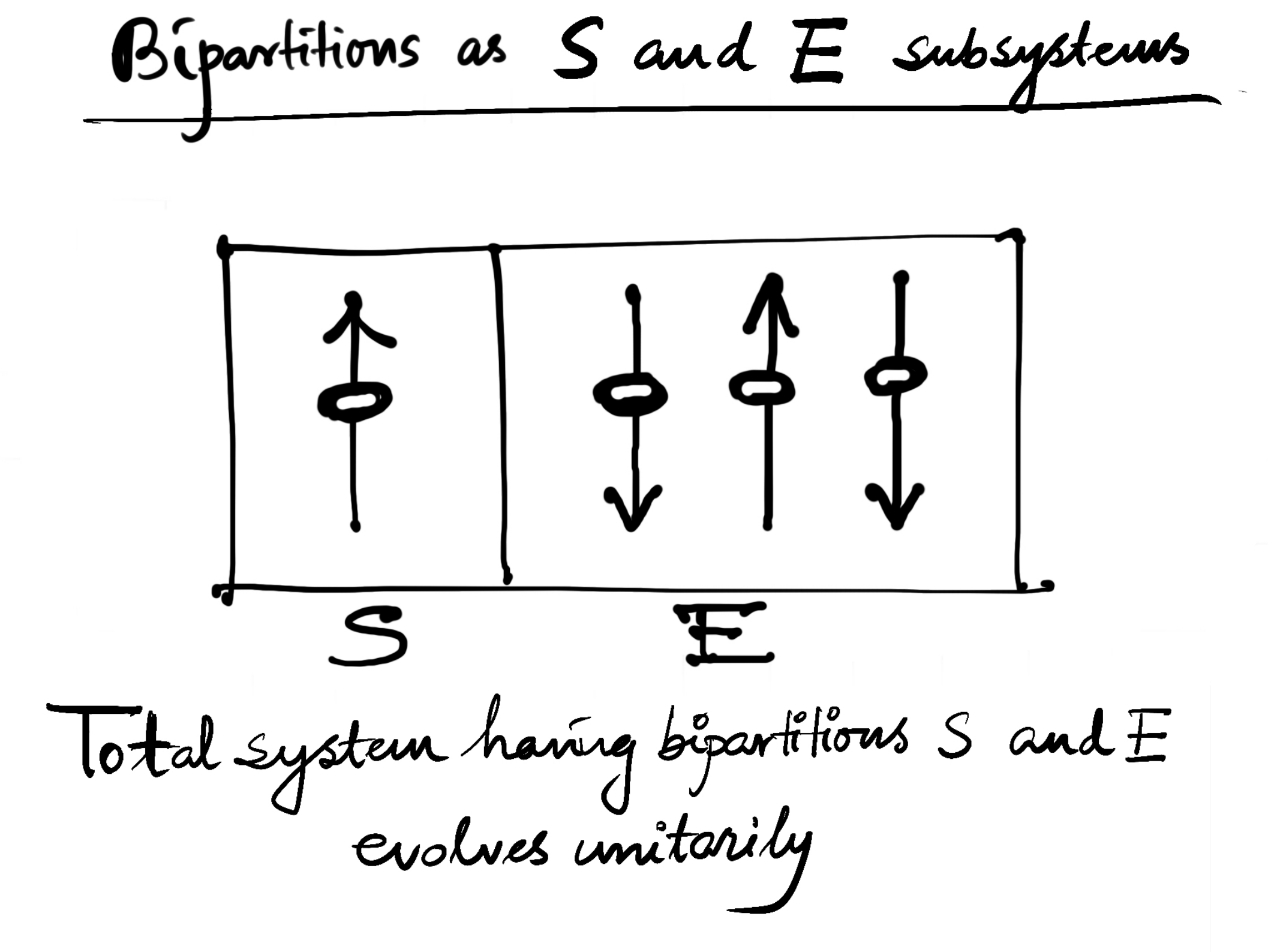}
    \caption{A sketch illustrating the bipartition of the Ising model as the system and the bath.}
    \label{fig:ABSE_bipartition}
\end{figure}

In the literature, it has been shown that entropy production has a natural connection to information scrambling \cite{zanardibipotoc2,Landientropyprod}. Further, the connection between the operator entanglement and bipartite OTOC was brought out in \cite{Zanardiopenbipotoc1}. To account for the correlation between the subsystems, a useful quantity, that is, correlation entropy ~\cite{EspositoEntprodentcorre}, is also studied. To study the interplay between these quantities, we take up a simple toy model of an Ising spin system in a tilted magnetic field, considered above, and analyze its global unitary evolution. Here, the partition $A$ is taken as the $S$ subsystem of 1 spin, and the partition $B$ of 3 spins is the $E$ subsystem, see Fig.~\ref{fig:ABSE_bipartition}. The $S$ could be thought of as a system, and $E$ as the bath. After illustrating the above ideas for this toy model, we take up the Dicke and TC models and study their entropy production and correlation entropy. 

The operator entanglement in many-body quantum systems quantifies how much the local operators become entangled over time. It serves as a natural candidate for distinguishing between chaotic and integrable dynamics in quantum systems~\cite{Medenjak}. Further, it was shown in~\cite{zanardibipotoc2,Zanardiopenbipotoc1,Zanardi2001opent} that the bipartite OTOC for a unitary evolution $U$ matches the operator entanglement $E_{op}(U)$.  The total unitary $\ket{U}_{ABA'B'} = U_{AB} \otimes I_{A'B'}\ket{\psi^+}$ for the maximally entangled state $\ket{\psi^+} = \frac{1}{\sqrt{d}}\sum_{j = 0}^{d-1}\ket{j}_{AB}\ket{j}_{A'B'}$ is supported over the composite Hilbert space $\mathcal{H}_{AB}$ and its replica $\mathcal{H}_{A'B'}$. Since the entanglement of state $\ket{U}_{ABA'B'}$ across $AA'|BB'$ is non-trivial, in contrast to the bipartition $AB|A'B'$ which is maximal due to it being local unitarily equivalent to the maximally entangled state, $B$ and $B'$ are traced out. The explicit expression for $E_{op}(U)$ is given by the linear entropy of $\sigma_U$ \cite{zanardibipotoc2,Zanardi2001opent}.
\begin{align}
    E_{op}(U)=1-{\rm Tr}(\sigma^2_U),
\end{align}
where
\begin{equation}
    \sigma_U={\rm Tr_{BB'}}\left(\ket{U}\bra{U}\right).
\end{equation}

The loss of distinguishability between the subsystems $S$ and $E$ motivates the study of entropy production $\Sigma$ that quantifies irreversibility emerging from discarding any information contained locally in the state of $E$, wherein the non-local information shared between $S$ and $E$ becomes essential. 
For the state of the system $\rho_S(t)$ given by 
\begin{align}
    \rho_S(t) = {\rm Tr}_E \left[\rho_{SE}(t)\right]={\rm Tr}_E\left[U\left\{\rho_S(0)\otimes\rho_E(0)\right\}U^\dagger\right],
\end{align}
where $U = e^{-iHt}$, the entropy production is~\cite{EspositoEntprodentcorre, Landientropyprod}
\begin{align}\label{entropy_production_general}
    \Sigma=S(\rho_{SE}(t)||\rho_S(t)\otimes \rho_E(0)),
\end{align}
where $S\left[(*)||(\circ)\right]={\rm Tr}\left[(*) \ln (*) - (*) \ln (\circ)\right]$ is the quantum relative entropy. 
It is noteworthy that the sum of the $\rho_S(t)$ and $\rho_E(t)$ entropies is not the entropy of the total system as it lacks the entropy contribution contained in the correlations, in particular entanglement between the two subsystems~\cite{EspositoEntprodentcorre}. To account for this, correlation entropy $S_C(t)$ is defined.
It accounts for the entropy contribution due to entanglement between the system and the reservoir and is given by
\begin{align}\label{correlation_entropy}
    S_C(t) =-S(\rho_{SE}(t)||\rho_S(t)\otimes \rho_E(t)).
\end{align}

For the unitary evolution, the bipartite OTOC, Eq.~\eqref{Bip otoc unitary}, shows the scrambling of information between the two subsystems with the evolution of time as the initial local behavior of the system gets altered. 
\begin{figure}
    \centering
    \includegraphics[width = 1\columnwidth]{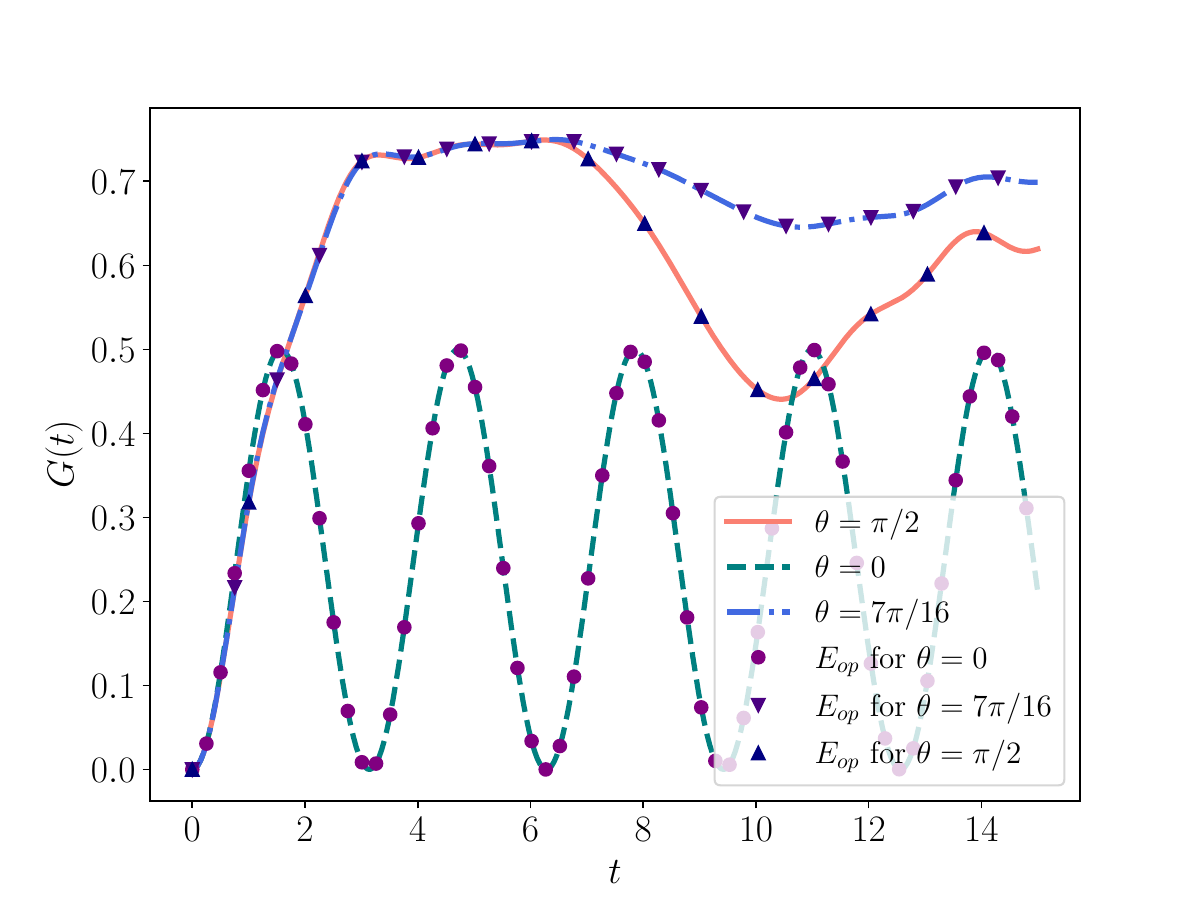}
    \caption{Comparison of the bipartite OTOC and operator entanglement $E_{op}(U)$ for three different angles, where the bipartition is outlined in Fig.~\ref{fig:ABSE_bipartition}, and the total system evolves unitarily. Here, $J=\mathcal{B}=0.5$.}
    \label{Bipartite_OTOC_and_operator_entanglement_ABIsing}
\end{figure}
As can be seen from Fig.~\ref{Bipartite_OTOC_and_operator_entanglement_ABIsing}, the bipartite OTOC increases initially and then varies around a saturation value, except for the case where $\theta = 0$ (integrable case). For $\theta= 0$, we observe an oscillatory behavior of the bipartite OTOC. The bipartite OTOC is higher for the angle $\theta = 7\pi/16$ when compared with the bipartite OTOC for $\theta = \pi/2$. The long-time behavior of the curve for $\theta= \pi/2$ shows deviation from the saturation value, consistent with the behavior observed in Fig.~\ref{fig:Theta_model_bip_OTOC_corrected}. Interestingly, the operator entanglement for all the angles matches exactly with the variation of the bipartite OTOC. This benchmarks the statement that the operator entanglement serves as a witness of (non-)integrable dynamics of quantum systems for unitary evolution. 
\begin{figure}
    \centering
    \includegraphics[width=1\columnwidth]{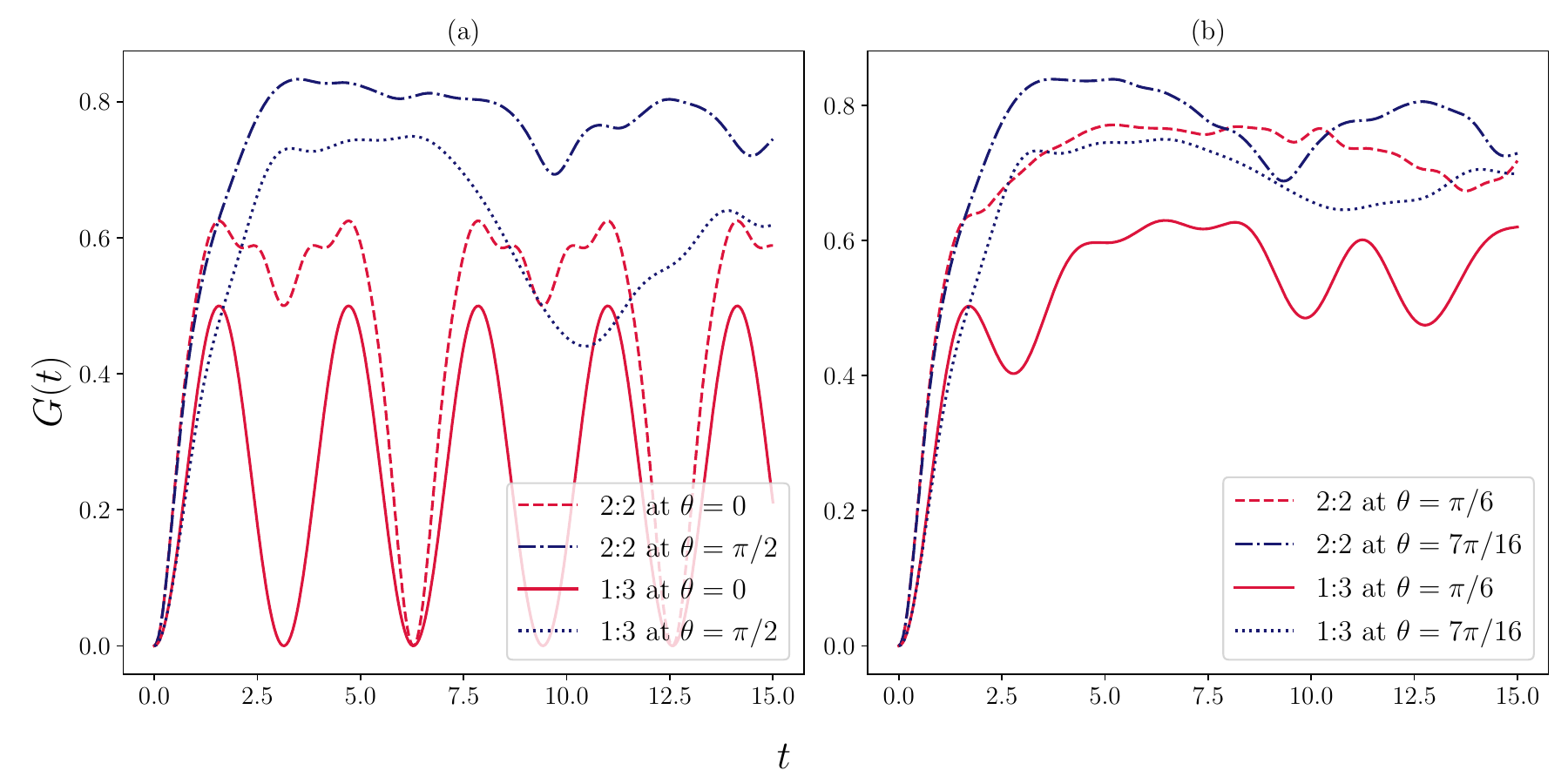}
    \caption{The bipartite OTOC for unitary evolution of a system of 4 spins having two different bipartitions: $A: B = 1:3$ and $2:2$. In (a), we take the case for the angle of tilt $\theta=0$ (trivially integrable) and $\theta=\pi/2$ (non-trivially integrable). In (b), we take two angles, $\theta=\pi/4$ and $\theta=7\pi/16$ (non-integrable). The parameters are taken as $J=\mathcal{B}=0.5$.}
    \label{fig:comparison_bipartition_ABIsing_bipOTOC}
\end{figure}
In Fig.~\ref{fig:comparison_bipartition_ABIsing_bipOTOC}, we compare the bipartite OTOC for two different bipartitions $A: B = 1:3$ and $2:2$ for the system of 4 spins, Fig.~\ref{fig:ABSE_bipartition}. We observe that the bipartite OTOC always takes higher values for the 2:2 bipartition case as compared to the 1:3 bipartition case. Further, the qualitative behavior of the bipartite OTOC is similar for $\theta = 0$ and $\pi/2$ for both kinds of bipartitions. This behavior is absent in the non-integrable regime.
\begin{figure}
    \centering
    \includegraphics[width = 1.0\columnwidth]{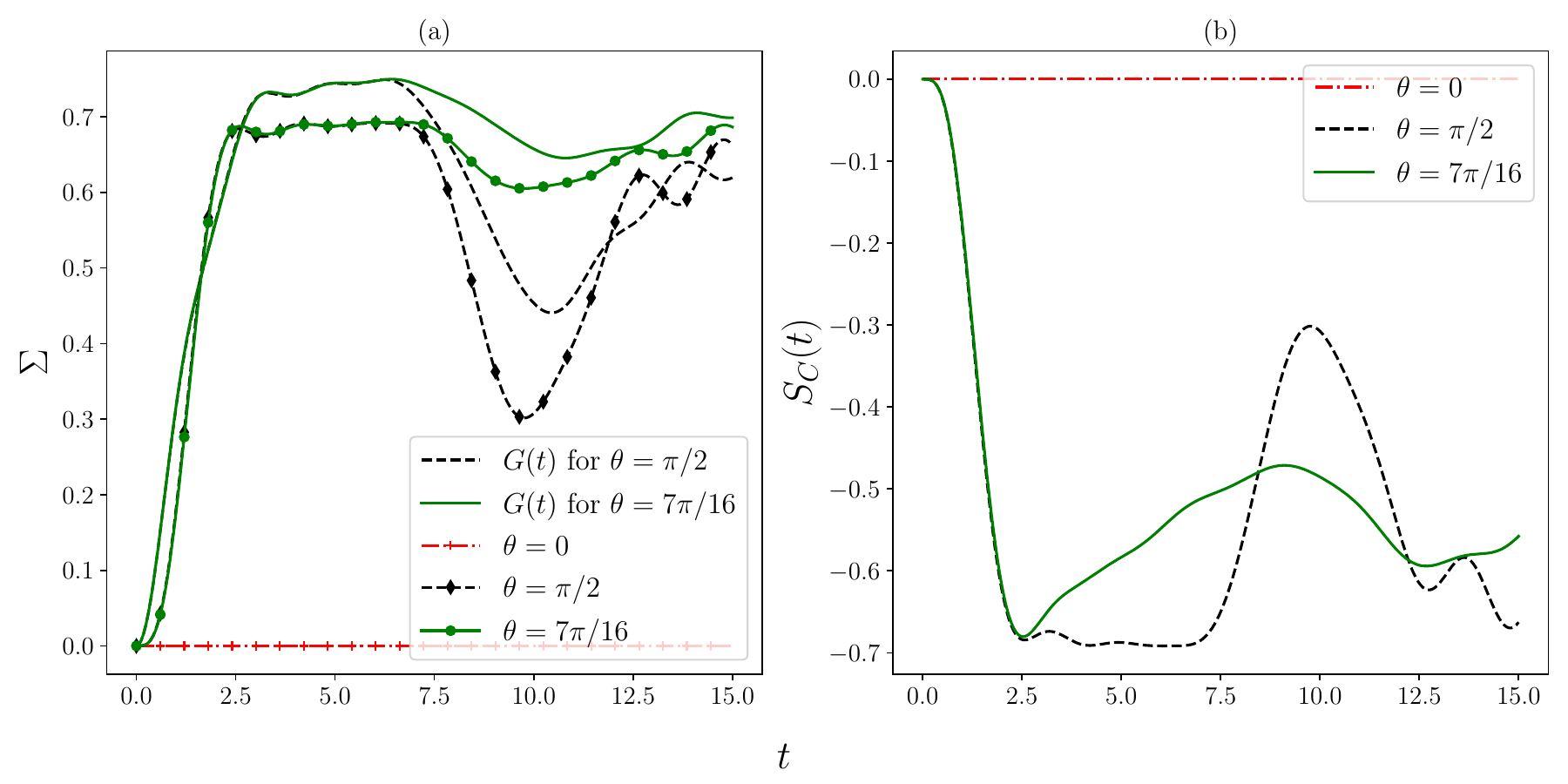}
    \caption{(a) Entropy production $\Sigma$ along with bipartite OTOC $G(t)$ and (b) correlation entropy $S_C$ of the $S$-$E$ Ising spin system for three different angles of tilt of the magnetic field. The parameters are $\mathcal{B}= J = 0.5$.}
    \label{fig:Ent_Int_Ent_corre_ABIsing}
\end{figure}%

The entropy production $\Sigma$ and the correlation entropy $S_C$ for the tilted Ising model are calculated using the ground state as the initial state for the subsystem $S$ and the maximally mixed state for the subsystem $E$. Both are plotted in Fig.~\ref{fig:Ent_Int_Ent_corre_ABIsing} along with the bipartite OTOC $G(t)$. It is observed that the entropy production $\Sigma$ rises quickly, reminiscent of the behavior of the bipartite OTOC, when the angle of tilt of the magnetic field is non-zero. The highs and the lows of the bipartite OTOC match those of the entropy production. 
The similarity of bipartite OTOC with entropy production hints that entanglement is the underlying factor that affects the commutativity of any two unitary operators in their respective subspaces and the entropy production. It is noteworthy that the bipartite OTOC reveals the changes through the operator evolution, and the entropy production does it through the state evolution. This is consistent with the observation made in~\cite{Deffner_2021}, where a connection between the bipartite OTOC and the mutual information between the subsystems was established. It can be pointed out here that the entropy production, Eq.~\eqref{entropy_production_general}, can also be written as the sum of the mutual information between the subsystems $S$ and $E$ and the Kullback Leibler divergence between the states $\rho_E(t)$ and $\rho_E(0)$. Therefore, analogous to the relationship between the bipartite OTOC and the mutual information, a similar connection between the entropy production and the bipartite OTOC can be established.
Interestingly, for some time periods, the entropy production rate is negative, which was used as an identifier of non-Markovianity in~\cite{Rivas_measure}. 
\begin{figure}
    \centering
    \includegraphics[width = 1\columnwidth]{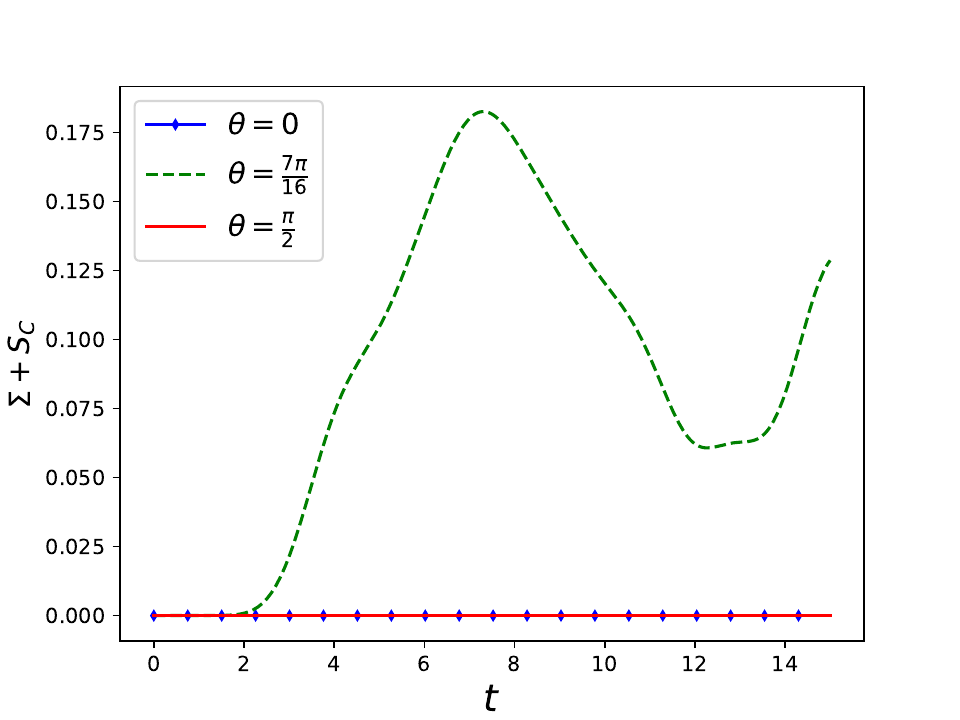}
    \caption{The quantity $\Sigma+S_C$ as a function of time for the tilted field Ising model. The parameters are $\mathcal{B} = J = 0.5$.}
    \label{fig:Ent_Int_ABIsing}
\end{figure}
Further, the correlation entropy $S_C$ is plotted in Fig.~\ref{fig:Ent_Int_Ent_corre_ABIsing}(b). It can be seen that the values of $|S_C|$ are lesser than the values of the entropy production, that is, $\Sigma + S_C \ge 0$. For $\theta = 0$, we observe from Figs.~\ref{fig:Ent_Int_Ent_corre_ABIsing}(a) and (b) that both the entropy production and the correlation entropy are zero, indicating its trivially integrable behavior. For $\theta = \pi/2$, the entropy production and the correlation entropy match exactly, modulo with opposite signs, that is, the sum $\Sigma + S_C = 0$ (Fig.~\ref{fig:Ent_Int_ABIsing}). Hence, the irreversibility due to the lack of non-local information shared between the two subsystems $S$ and $E$ is exactly compensated by the entropy contribution due to entanglement between them. This indicates that the Ising model is non-trivially integrable at the angle $\theta = \pi/2$; in fact, it can be made integrable by mapping it to a model of non-interacting fermions using the Jordan-Wigner transformation. There's a mismatch between $\Sigma$ and $S_C$ for the angle $\theta = 7\pi/16$, which is evident from Figs.~\ref{fig:Ent_Int_Ent_corre_ABIsing} and \ref{fig:Ent_Int_ABIsing}, indicating a non-integrable behavior of the system. Interestingly, these features of the Ising model are corroborated here by tools of quantum statistical mechanics, that is, entropy production and correlation entropy. We note here that in order to facilitate a comparison with bipartite OTOC, the entropy production and correlation entropy are calculated by taking the $E$ subsystem in the maximally mixed state and the $S$ subsystem in any state, taken here as the ground state. This choice is followed subsequently. The choice of different states affects the difference in values between the bipartite OTOC and the entropy production, even though the qualitative behavior remains similar. 
\begin{figure}
    \centering
    \includegraphics[width=1\columnwidth]{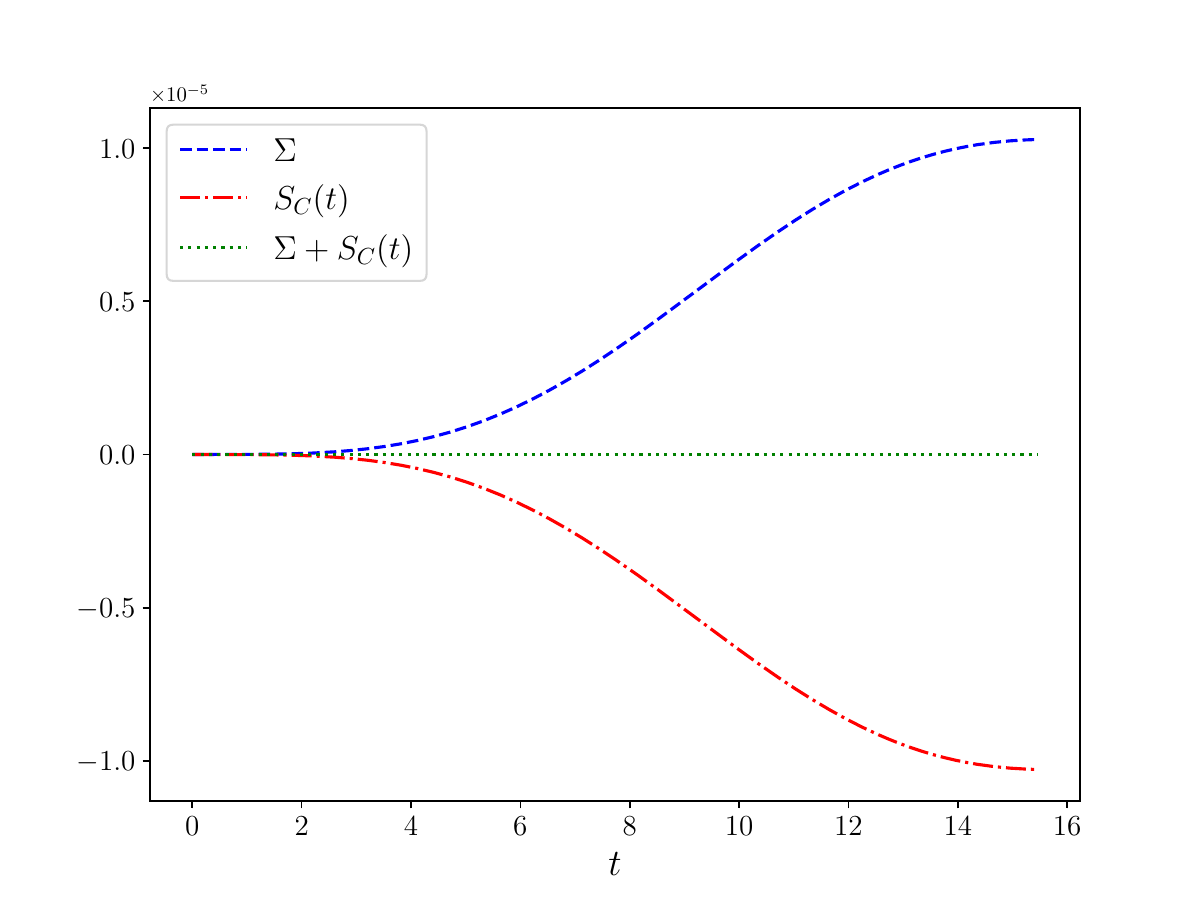}
    \caption{Variation of the entropy production for the tilted field Ising model, Fig.~\ref{fig:ABSE_bipartition}, in the weak $S$ and $E$ coupling scenario with the number of spins increased to 6 in the $E$ subsystems. The values of $\mathcal{B} = J = 0.5$ and $\theta = \pi/2$. Further, interaction strength between the spins 1 and 2 is taken to be $10^{-3}J$.}
    \label{fig:Entropy_production_Ising_model_GKSL_approximation}
\end{figure}
Also, we plot the variation of the entropy production and the correlation entropy for the tilted field Ising model with weak $S$ and $E$ coupling strength for the angle $\theta = \pi/2$ in Fig.~\ref{fig:Entropy_production_Ising_model_GKSL_approximation}. This is done to simulate the impact of the GKSL type of evolution on the entropy production. It is observed that the values of entropy production are very low (of the order of $10^{-5}$) as compared to the previous cases, and they increase as time progresses. Again, in this case of $\theta = \pi/2$, the sum of the entropy production and the correlation entropy is zero.
\begin{figure}
    \centering
    \includegraphics[width = 1\columnwidth]{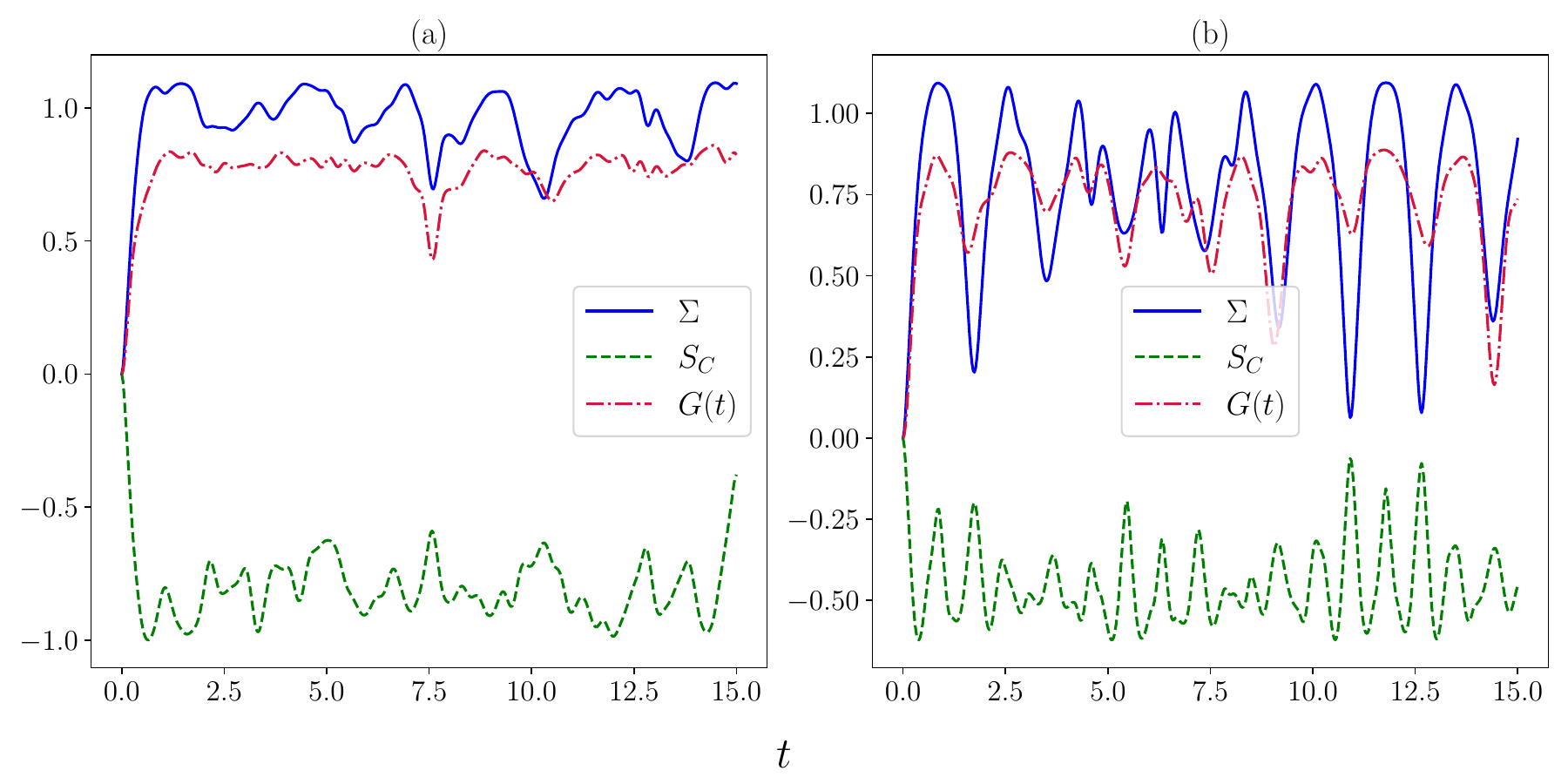}
    \caption{Variation of the entropy production $\Sigma$, correlation entropy $S_C$, and bipartite OTOC $G(t)$ with time for the (a) Dicke and (b) TC models, under unitary evolution. Here, (a) $\lambda = 1.5$, and (b) $\lambda = 2$, and $\omega_c = \omega_0 = 2$.}
    \label{bip_OTOC_entropy_production_Dicke_TC}
\end{figure}%

To further motivate the connection between entropy production and bipartite OTOC, we briefly return to the Dicke and TC models considered above. Figure~\ref{bip_OTOC_entropy_production_Dicke_TC} depicts the entropy production, correlation entropy, and bipartite OTOC for the Dicke and the TC model undergoing unitary evolution. Here, the atomic and the radiation field parts are the subsystems $A$ and $B$, respectively, for the calculation of the bipartite OTOC and are analogous to the symbols $S$ and $E$ for the calculation of the entropy production, Eq.~\eqref{entropy_production_general}, and correlation entropy, Eq.~\eqref{correlation_entropy}. Here, the subsystems $S$ and $E$ are initially taken to be in the ground state and the maximally mixed state, respectively.
It can be observed that the variations in the entropy production and the bipartite OTOC are in good agreement with each other. 
This suggests a relationship between irreversibility, characterized by entropy production, and information scrambling, characterized by bipartite OTOC, of the system. However, the numerical values of the two do not coincide, indicating that information scrambling and irreversibility, though similar, are not synonymous. Further, in both cases, the values of the correlation entropy are lesser than the entropy production, and in the case of the Dicke model, $S_C \sim -\Sigma$. 

The features brought out above are relevant under unitary dynamics. To explore the scenario of open quantum systems, we calculate the entropy production and bipartite OTOC for the Dicke model under the influence of the GKLS master equation. The entropy production, Eq.~\ref{entropy_production_general}, simplifies to the following form in the weak coupling limit invoked in the GKLS regime~\cite{EspositoEntprodentcorre} 
\begin{align}
    \Sigma = S(\rho_S(0)||\rho_S^{eq}) - S(\rho_S(t)||\rho_S^{eq}),
\end{align}
where $\rho_S^{eq} = e^{-\beta H_{Dicke}}/Z$, with $Z = {\rm Tr} \left[e^{-\beta H_{Dicke}}\right]$. 
\begin{figure}
    \centering
    \includegraphics[width=1\columnwidth]{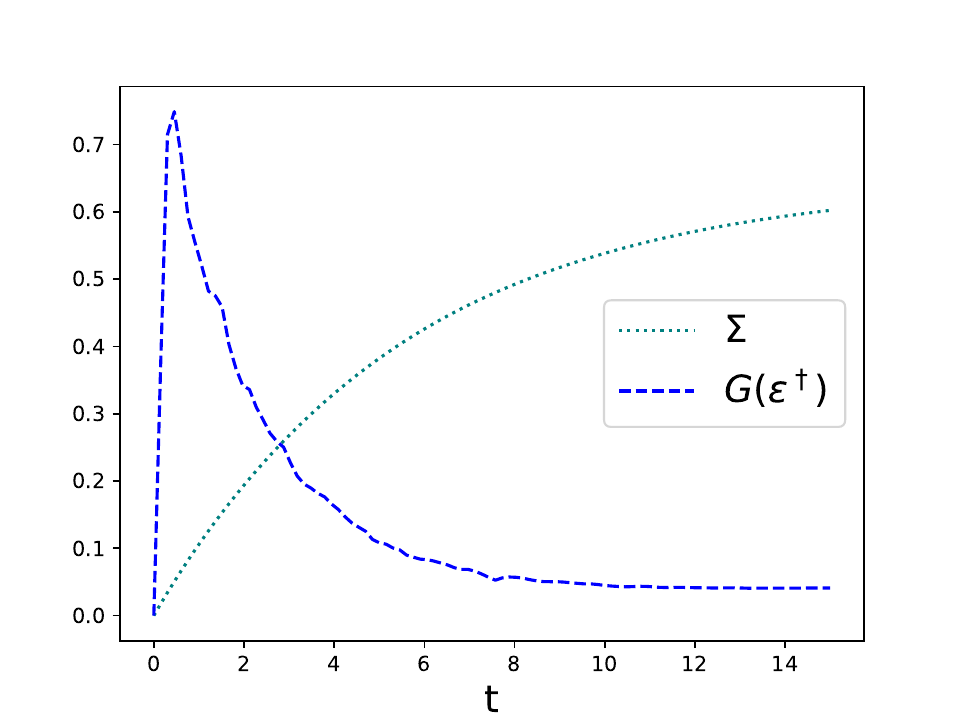}
    \caption{Entropy production and bipartite OTOC for the Dicke model at $\lambda=1.5$, $T=2$ and $\omega_c = \omega_0 = 2$, $\gamma=\kappa=0.05$.}
    \label{Dicke_entropy_bip_otoc}
\end{figure}
As can be seen from Fig.~\ref{Dicke_entropy_bip_otoc}, the entropy production is non-negative and also monotonic, consistent with its inherent Markovian behavior. Also plotted along with it is the bipartite OTOC. In contrast to the above case of unitary evolution, the connection between bipartite OTOC and entropy production is not evident here. The bipartite OTOC, in this case, dissipates with time after the initial rise, whereas the entropy production keeps on increasing till eventual saturation is achieved (not depicted here).

\section{Conclusion}\label{sec_conclusion}
In this work, a study of information scrambling using bipartite OTOC was made. The bipartite OTOC provides a general framework wherein any specific choice of unitary operator is eliminated. The bipartite OTOC of the atom-field interactions models, namely, the Dicke and TC models and the Ising model with a tilted magnetic field, were studied. In the context of open quantum systems, the operators were evolved via the adjoint form of the GKLS master equation. This revealed the regions where information scrambling occurred. 

To bring out the connection between information scrambling and irreversibility, bipartite OTOC was examined in conjunction with entropy production and correlation entropy. For this, we used the simplified forms of the Ising model with a tilted magnetic field, as well as the Dicke and TC models. Interestingly, a qualitative relationship between entropy production, characterizing irreversibility, and bipartite OTOC, characterizing information scrambling, emerged from this under unitary dynamics. This was, however, not discernible in the case of open quantum systems. The exact match of the bipartite OTOC and operator entanglement for all angles of tilt was observed for the Ising model. Moreover, the (non-)integrable features of the Ising model with a tilted magnetic field were validated using the tools of quantum statistical mechanics, that is, entropy production and correlation entropy. For the angle of tilt $\theta = 0$, entropy production and correlation entropy were zero, while their sum was zero for the angle $\theta = \pi/2$, suggesting a trivial integrable nature of the model at the former angle, and non-trivial integrability in the latter case. For any intermediate angle, say $\theta = 7\pi/16$, the sum was greater than zero, suggesting non-integrability. 

\appendix

\section{The swap operators}
To derive the quantity $G(t)$, the following identity involving Haar integral over random unitary matrices is used. 
\begin{align}
    \left( \int_ {U{(d)}}  U \otimes U^{\dagger}  d\mu(U) \right) = \frac{S}{d}.
\end{align}
That's how we are introduced with the swap operator $S$. To elaborate on the structure of the swap operator,
let us introduce another replica of the original Hilbert space, $\mathcal{H}^{\prime}=\mathcal{H}_{A^{\prime}} \otimes \mathcal{H}_{B^{\prime}}$ with the same dimension as that of $\mathcal{H}$, i.e., $d={\rm dim}(\mathcal{H})=d_Ad_B$.
Taking $S_{AA^{\prime}}$ an operator over $\mathcal{H} \otimes \mathcal{H^{\prime}}$, we get 
\begin{equation}\label{eq:5}
    G(t)=1-\frac{1}{d^2}{\rm Tr}(S_{AA^{\prime}}U_t^{\otimes2}S_{AA^{\prime}}U_t^{\dagger \otimes2}),
\end{equation}
the same expression is valid for $S_{BB^{\prime}}$ also.
The following are the structure of $S$, $S_{AA^{\prime}}$ and $S_{BB^{\prime}}$
\begin{align}
    S=\sum_{a,b,a^{\prime},b^{\prime}}\ket{aba^{\prime}b^{\prime}}\bra{a^{\prime} b^{\prime} a b }, \\
    S_{AA^{\prime}}= \sum_{a,b,a^{\prime},b^{\prime}}\ket{aba^{\prime}b^{\prime}}\bra{a^{\prime} b a b^{\prime} }, \\
    S_{BB^{\prime}}=\sum_{a,b,a^{\prime},b^{\prime}}\ket{aba^{\prime}b^{\prime}}\bra{a b^{\prime} a^{\prime} b }.
\end{align}
Here, all of the swap operators belong to $\mathcal{H}\otimes \mathcal{H^{\prime}}$ or more precisely in a $A+B$ bipartite system, to  $\mathcal{H}_A \otimes \mathcal{H}_B \otimes 
  \mathcal{H}_A ^{\prime} \otimes \mathcal{H}_B^{\prime} $.

\section{Calculation of the dynamical map and the swap operators}
\subsection{The dynamical map}
To calculate the bipartite OTOC, the adjoint map $\mathcal{E}^\dagger$ and swap operator $S_{AA'}$ are needed. For the GKSL assumptions, we have the following relation between the dynamical map and the Lindbladian superoperator,
\begin{align}
    \mathcal{E}^\dagger (t) &= e^{\mathcal{L}^\dagger t}.
\end{align}
Now, vectorizing the GKSL master equation, we obtain $\mathcal{L}^\dagger$.
\begin{align}
    \mathcal{L}^\dagger&=\left(H_m\otimes \mathbb{I}_d-\mathbb{I}_d\otimes H_m^{T}\right) + \sum_{k=0}^{N} \gamma_k \left\{ \left( \hat{L}_k^T \otimes \hat{L}_k^* \right) \right.\nonumber\\
    & - \left.\frac{1}{2}\left(\hat{L}_k^{\dagger} \hat{L}_k \otimes \mathbb{I}_d +  \mathbb{I}_d \otimes (\hat{L}_k^{\dagger} \hat{L}_k)^T \right) \right\}.
\end{align}
So, the map $\mathcal{E}^\dagger(t)$ becomes a $d^2 \times d^2$ dimensional matrix after exponentiating the above Lindbladian superoperator matrix, where all the Hamiltonians and Lindblad operators were $d \times d$ dimensional before vectorizing. Here $d=d_Ad_B$, where $d_A$ and $d_B$ are the dimensions of partitions $A$ and $B$, respectively.
In our work, $H_m$ stands for three different hamiltonians for $H_{Dicke}\text{, }H_{TC}\text{, }H_{\theta}$. The corresponding Lindblad operators for these models are described in the text. In the case of the Dicke and the TC models, we chose $d_A=3$ and $d_B=3$ for our calculations.
The partitions for different scenarios for the TFIM model are discussed in the text. For $k$ spins in the $A$ subsystem, we have $d_A=2^k$, and for $n$ spins in the $B$ subsystems, we have $d_B=2^n$. 
\subsection{The specific swap operators}
The swap operator $S_{AA'}$ for the Dicke, TC, and the tilted field Ising model is
\begin{align}
    S_{AA'}^{Dicke/TC/TFIM}&=\sum_{j=0, k=0}^{d_A-1, d_A-1} \ket{j}\bra{k}_A\otimes\mathbb{I}_B\otimes\ket{k}\bra{j}_{A^\prime}\otimes\mathbb{I}_{B^\prime},
\end{align}%
where the dimension $d_A$ in each case is described above.
We vectorize ($\text{vec}\left\{\sum_{i, j} \ket{i}\bra{j}\right\} = \sum_{i, j}\ket{ij}$) the above swap operator and apply the map $\mathcal{E}^\dagger\otimes\mathcal{E}^\dagger$ on it. This is then de-vectorized and fed to Eq.~\eqref{eq_avg_bip_otoc_open} to calculate the average bipartite OTOC.

\bibliography{reference}
\bibliographystyle{apsrev}

\end{document}